# Current induced magnetization switching in exchange biased spin-valves for CPP-GMR heads

A. Deac[1,3*], K.J. Lee[1,3], Y. Liu[2], O. Redon[1,3], M. Li[2], P. Wang[2], J.P. Nozieres[1], B. Dieny[1]

1. SPINTEC, URA CEA/CNRS, CEA Grenoble/DRFMC, 17, Rue des Martyrs, 38054 Grenoble cedex 9, France
2. Headway, 678 Hillview Dr, Milpitas CA 95035
3. CEA Grenoble / DRT / Leti, 17, Rue des Martyrs, 38054 Grenoble cedex 9, France

* alina.deac@cea.fr

## Abstract

In contrast to earlier studies performed on simple Co/Cu/Co sandwiches, we have investigated spin transfer effects in complex spin-valve pillars with a diameter of 130nm developed for current-perpendicular to the plane (CPP) magneto-resistive heads. The structure of the samples included an exchange biased synthetic pinned layer and a free layer both laminated by insertion of several ultrathin Cu layers. Despite the small thickness of the polarizing layer, our results show that the free layer can be switched between the parallel (P) and the antiparallel (AP) states by applying current densities of the order of $10^7$ A/cm^2. A strong asymmetry is observed between the two critical currents $I_c^{AP-P}$ and $I_c^{P-AP}$, as predicted by the model of Slonczewski model. Thanks to the use of exchange biased structures, the stability phase diagrams could be obtained in the four quadrants of the (H, I) plan. The critical lines derived from the magnetoresistance curves measured with different sense currents, and from the resistance versus current curves measured for different applied fields, match each other very well. The main features of the phase diagrams can be reproduced by investigating the stability of the solutions of the Landau Lifshitz Gilbert equation including spin torque term within a macrospin model. A spin-transfer saturation effect was observed in the positive currents range. We attribute it to a de-depolarization effect which appears as a consequence of the asymmetric heating of the pillars, whose top and the bottom leads are made of different materials.

## Introduction

It was predicted[1, 2] that a spin-polarized electrical current flowing through a magnetic layer can exert a large torque on its magnetization, inducing magnetic excitations and possibly the switching of its magnetization. The first experimental results demonstrating that it is indeed possible to switch the magnetization of a layer back and forth by applying a spin-polarized current[3] have been received with considerable interest. Motivated by its potential application as an alternative write scheme in magnetic random access memories (MRAM) or in magnetic recording technology, many efforts have been made ever since in order to understand the physics of this new phenomenon: new theoretical models[4-6] have been published, as well as numerical simulations[7-10] and experimental results[11-16]. At an earlier stage, the effect was studied only in very simple structures of the type *magnetic thick layer / non-magnetic spacer / magnetic thin layer*, where the two magnetic layers were either CoFe or NiFe. Lately, however, more complicated structures were investigated, including CPP spin-valves with pinned synthetic layers[17-19] and even magnetic tunnel junctions (MTJ)[20]. More interest in this field has been arisen by recent experiments showing that a spin-polarized current can drive the magnetization of a layer into steady precessional modes inaccessible by applying only a magnetic field[21, 22]. These effects can be applied in new magnetic devices, such as resonators or microwave sources.

On the other hand, nowadays, the trend in high density magnetic recording technology is to replace the current-in-plane (CIP) geometry in magneto-resistive heads for computer disk drives by the CPP configuration, since the latter offers larger magneto-resistance (MR) ratios and allows for higher storage density by reducing the shield to shield spacing. The required current densities are between $10^7$ A/cm² and $10^8$ A/cm², of the same order of magnitude as the currents at which spin-torque induced magnetic excitations are observed. Such effects can generate noise and influence the biasing of the magnetic heads; it is therefore important to study and understand them in order to control their influence.

## Sample preparation and experimental set-up

The present experiments were conducted on sputtered spin-valves with the following structure: *IrMn 7/ AP2 4.0 / Ru 0.8/ AP1 4.4/ Cu 2.6/ F 3.6/ Ta*. All thicknesses are given in nm (see Fig. 1). The bottom electrode is a 1μm thick NiFe stripe, with its longer dimension (12μm) along the pinning direction of the pinned layer; the top electrode is patterned from a Cu layer. The NiFe stripe is meant to constitute one of the magnetic shields in the real device. AP1 (polarizing) and F (free) are CoFe layers, laminated by the insertion of three ultra-thin (0.3nm) Cu layers. The purpose of the lamination is to increase the resistance of the part of the spin-valve which is active from the point of view of the CPP GMR, i.e. the AP1/Cu/F sandwich. Indeed, at the lowest order of approximation, this active part can be considered as connected in series with the other layers which are important to insure suitable magnetic properties for read head applications but reduce the CPP GMR ratio because of their additional serial resistance. The increase in resistance due to lamination is significant since each CoFe/Cu interface has a resistance equivalent to about 4nm of bulk CoFe layer[23]. An enhancement from 1.5% to 2.2% of the CPP-GMR was observed in these structures due to the lamination[27]. However, this resulting increase in CPP-GMR amplitude is lower than expected from the relative increase of resistance because of significant spin-flip at each CoFe/Cu interface. In Ref.24, it was indeed shown that the conduction electrons loose about 25% of their polarization at each Co/Cu interface. The same order of magnitude of depolarization may be expected at $Co_{50}Fe_{50}$/Cu interface. As a consequence, the electrons are almost fully repolarized along the direction of the local magnetization after having traversed less than 2 nm of the laminated stacks. This quite short effective spin diffusion length in the laminated layers is responsible for the moderate benefit of lamination.

After the patterning of the bottom lead, electron-beam lithography and ion-beam etching were used to fabricate square pillars (with rounded corners) with a lateral size of 130nm. An insulator layer (alumina) was then deposited. The top-contacts were opened through a lift-off process; the last steps were the sputtering and the patterning of the top Cu lead, perpendicular to the bottom one. The layout allowed for four-probe point measurements, with two contacts placed on the top Cu lead and two on the bottom NiFe lead.

A simple set-up was used for the experiments. External magnetic fields up to +/-600 Oe could be applied using a small electromagnet, and a Keithley 2400 source meter was used both as a current source and as a voltmeter. Although quite reliable as a current source, the Keithley 2400 is less accurate as a voltmeter, which explains why sometimes a slight divergence was observed in the resistance versus current (R(I)) curves around I = 0. The resistance of the samples ranged between 5.6 and 8.8 Ω and the magnetoresistance amplitude was around 2.2%. The coercivity of the free layer ranged between 10 and 150 Oe, i.e. our soft layer was much softer than the thin layers used in all previous experiments reported in the literature. The dispersion from sample to sample was probably due to small differences in the detailed shape of the pillars, especially at their edges. A shift of a few tens of Oe was measured at low current in most samples in the position of the minor hysteresis loop associated with the switching of the free layer. This shift is due to the magnetostatic stray field from the pinned layer. As a result, in zero applied magnetic field, the samples were in the antiparallel state. This means that the stray field corresponds to a dominant interaction with the AP1 layer which is closer to the free layer and slightly thicker than the AP2 layer.

Alternatively, magnetoresistance curves could be measured using a KLA Tencor tester with a maximum available field of +/-1200 Oe. In most of the samples, the magnetization of the pinned layer started to switch around 1000 Oe, meaning that in the range +/-600 Oe, it remained unaffected. The KLA Tencor tester offers also the possibility of measuring the MR properties at temperatures ranging from room temperature (25°C) to 110°C. This option was used in order to determine the thermal variation of the resistance in the two magnetic configurations (parallel (P) and antiparallel (AP)) (Fig. 5).

All the other experiments were conducted at room temperature.

**Results**

In the discussion of the results, we use the following conventions (Fig. 1):
1) Negative magnetic field is oriented along the magnetization of the pinned AP1 layer; therefore, it favors the parallel alignment of the magnetizations of the two layers. (It follows that positive field favors the antiparallel orientation.);
2) For negative current, the electrons flow from the free to the pinned layer, favoring the antiparallel state. (Inversely, for positive current, the electrons move in the opposite direction and favor the parallel state).

Considering the size of our samples, an applied current of 1mA corresponds to a current density of $0.59*10^7$ A/cm².

Fig. 2a shows a minor MR loop measured with a sense current of - 0.400mA. At such current density (- $2.36*10^6$ A/cm²), we do not expect any spin-transfer induced effects. The coercivity of this sample is $H_c = 91$ Oe. The magnetostatic field from the synthetic pinned layer shifts the loop towards negative fields (i.e. the magnetostatic stray field is positive: $H_{ms}$= 48 Oe), thus favoring the AP state. The low resistance state, $R_{min}$ = 8.78Ω, corresponds to the P alignment, and the high resistance state, $R_{max}$ = 8.97Ω, to the AP configuration. The MR amplitude is 2.16%. The same relative resistance variation is found between the two resistance levels on the R(I) curve in fig. 2b; the values of the resistance in the two states are also very close to the ones measured in the R(H) loop ($R_{min}$ = 8.79Ω and $R_{max}$ = 8.99Ω). We can therefore conclude that we have observed current induced magnetization switching of the magnetization of the free layer between P and AP configurations. Starting with the sample in the AP state, a positive current $I_c^{AP-P}$= 2mA ($j_c^{AP-P}$ = 1.18*10$^7$A/cm²) is needed in order to switch to the P state. Increasing the current even more leads to the heating of the sample, as indicated by the parabolic increase in the sample resistance. When sweeping the current backwards, towards negative values, a P-AP transition occurs for $I_c^{P-AP}$ = -3.3mA ($j_c^{P-AP}$ = 1.95*10$^7$A/cm²), after which the sample remains in the AP state until a positive current is applied. This is in agreement with our convention regarding the sign of the current. The order of magnitude of the critical currents is the same as the other values so far reported in the literature. Both for the magnetoresistance and for the resistance versus current curves, the transitions between the two states are very sharp, indicating that sample is switching between two single-domain states.

In order to compare the R(I) and R(H) data, we have constructed the phase diagram characterizing the magnetic stability of the system in two different ways : For the same samples, we have measured the R(I) curves for different applied fields and the magnetoresistance loops R(H) for different sense currents.

Fig. 3 shows the evolution of the resistance versus current characteristics while increasing the applied negative magnetic field. Considering the conventions for positive fields (which favor the P state), and for the sign of the current (positive current favors the P alignment), as long as the external field is not large enough, the AP-P transition occurs in negative currents, and the P-AP transition is induced by positive currents, as expected. At -51 Oe, when the magnetostatic field from the pinned layer is approximately compensated by the external field, a strong asymmetry is observed between the two switching currents, as predicted by Slonczewski's ballistic model (1996)[2]. Increasing the external field induces a shift of the loop towards more negative currents. The AP-P transition shifts slowly at low fields (between 0 and approximately -140 Oe), and faster at larger fields. Simultaneously, the coercivity is gradually reduced, the AP-P transition being shifted more than the P-AP. At -227 Oe, the curve is practically reversible. At -315 Oe, the maximum applied current is no longer sufficient for inducing a P-AP transition, and the sample remains in the P state (under the influence of the applied field).

The behavior of the sample in positive applied fields is unusual (Fig. 4). When H = 0 Oe, a strong telegraph noise is measured in the range of currents where the sample should be in the P state (between $I_c^{AP-P}$ = 2.2mA and the maximum applied current, 8mA). This noise diminishes when increasing the field, but, at the same time, a gradual reversible transition towards a higher resistance state appears for high values of the current. This reversible transition is moving towards lower values of the current when the field is increased. Simultaneously, the AP-P transition induced by the spin-transfer is moving towards higher values of the current. For H > 37 Oe, the sample remains in the AP state and no switching is observed. The grey curve is measured at 600 Oe, when the sample is in the AP state.

Both Fig. 3 and Fig. 4. show raw data. The resistance change due to heating has not been subtracted, and it is highly asymmetric for the two directions of the current. This is a consequence of the different composition of the top and the bottom electrode (Cu and NiFe, respectively), known as the Peltier effect. When a voltage is applied on the junction between two metals, it induces a temperature gradient between the two leads. The sign of the temperature gradient depends on the sign of the voltage. In our case, for a positive voltage, the hot electrode is the Cu/pillar system; for a negative voltage, the hot electrode is the NiFe lead. The measured resistance variation is due to the combination of the Joule heating (~$I^2$) and the Peltier heating/cooling (~I). In order to estimate the variation in temperature due to the combination of Joule and Peltier effects, we have measured the thermal variation of the resistance of the sample in the range 25°C-110°C (see Fig.5). Fig.2b shows that increasing the current from 0.4mA to 8mA yields a resistance increase of 2.95%. The comparison with Fig.5 indicates that this corresponds to a raise in temperature of about 50°C at 8mA. For negative current, a decrease of resistance of about 0.5% is first observed from –0.4mA to –5mA (the Peltier cooling dominating the Joule heating) followed by an increase of 0.5% between –4mA and –8mA (the Joule heating dominating the Peltier cooling) (Fig.2b). According to Fig.5, this corresponds to a decrease then increase of temperature of less than 10°C. We underline that these estimations of temperature variations are averaged over the entire pillar. Locally, the temperature variation can be even larger. This interpretation in terms of Joule and Peltier effects is supported by the observation that samples having two identical leads do not show any heating dependence on the polarity of the current.

An alternative procedure for studying spin-transfer induced effects consists in measuring magnetoresistance curves for different applied currents (Fig. 6 and Fig. 7). Increasing the negative sense current (Fig. 6) up to –3mA induces a slight shift of the loop towards positive fields; this observation is in good agreement with the fact that for negative currents the spin transfer torque tends to stabilize the AP state, since the electrons are flowing from the free layer to the pinned layer. The coercivity is not much affected in this range of current. Between –3mA and –4mA, the P-AP transition jumps from ~0 Oe to ~200 Oe, and the coercivity is virtually zero. This is probably because at this value, the current density is large enough to induce the P-AP transition of the free layer. The AP-P transitions still occur under the influence of the applied magnetic field. Increasing the current over -4mA causes a faster shift of the loop, and probably the formation of a vortex distortion. The transition between the two states becomes more and more slanted. In addition, the magnetoresistance amplitude drops from 2.16% for I = -0.4mA to less than 1.5% at +/-7.5mA. When applying a positive sense current (Fig. 7) up to 2mA, the transition AP-P is shifting towards more positive values, while the P-AP transition remains practically unchanged. At I = 2mA, the switching becomes reversible. Further increasing the current only yields a more pronounced slanting of the transition, but no additional shift is measured.

A remarkably good agreement is obtained when superposing on the same plot the phase diagrams from the resistance versus current curves for constant applied (negative) magnetic fields, and from the magnetoresistance curves measured with different (negative and positive) sense currents (Fig. 8). Following the approach of Grollier et al.[11], we have plotted the switching currents ($I_c^{AP-P}$ and $I_c^{P-AP}$) or the switching fields ($H^{AP-P}$ and $H^{P-AP}$) if the loop was hysteretic, and the beginning and the end of the transitions ($I^{start}$ and $I^{end}$, or $H^{start}$ and $H^{end}$), if they were reversible. Four distinct regions are identified:
1) Both the P and the AP states are stable in-between the $I_c^{AP-P}$ and $I_c^{P-AP}$, respectively $H^{AP-P}$ and $H^{P-AP}$ curves;
2) Only the P state is stable above the $I_c^{AP-P}$/ $H^{AP-P}$ and $I^{start}$/ $H^{start}$ lines;
3) Only the AP state is stable under the $I_c^{P-AP}$/ $H^{P-AP}$ and $I^{end}$/ $H^{end}$ lines;
4) Neither state is stable between the $I^{start}$ and $I^{end}$, respectively $H^{start}$ and $H^{end}$ curves.

Both series of measurements (R(I) for different H and R(H) for different I) have been repeated, yielding similar results.

**Discussion**

Following the macrospin approach of the spin torque induced dynamics[25,27], and taking into account our conventions for the direction of the field and of the current, the Landau-Lifshitz-Gilbert equation of motion of the free layer magnetization can be written:

$$\frac{\partial \vec{m}}{\partial t} = -\gamma\, \vec{m} \times [H_{res}\vec{u}_x - H_d(\vec{m}\cdot\vec{u}_z)\cdot\vec{u}_z] + \alpha\cdot\vec{m}\times\frac{\partial \vec{m}}{\partial t} - \frac{\hbar}{2e}\cdot\frac{\gamma}{M_s A t}\cdot g(\theta)\cdot I \cdot \vec{m}\times[\vec{m}\times(-\vec{u}_x)]$$

(1)

where: $\vec{m}$ is the unit vector of the magnetization of the free layer;

$\gamma_0 = 1.76*10^{-11} s^{-1} T^{-1}$ is the gyromagnetic ratio;

$H_{res} = H + H_{ms} \mp H_c$ is the total field on the free layer, when its magnetization is close to the P (-) or AP (+) state;

$H$ is the external applied magnetic field, $H_{ms}$ the magnetostatic field from the pinned layer, and $H_c$ the coercivity of the free layer;

$H_d = 4\pi M_s$ is the demagnetizing field;

$\alpha$ is the damping constant;

I is the applied current;

$t$ the thickness, A the area of the free layer (in our case, 3.6nm);

$e = 1.6*10^{-19} C$ *is the* electron charge;

$\hbar$ = Plank's constant;

$g(\theta)$ is a function which describes the angular dependence of the spin torque ($\theta$ being the angle between the magnetization of the two layers;

$\vec{u}_x$ is the unit vector parallel to the magnetization of the free (and pinned) layer, $\vec{u}_z$ is the unit vector perpendicular to the plane of the layers (along the direction of the current) (see Fig. 1 for the definition of the coordinate system).

The first term in Eq. (1) is the field induced precession term; the second one is the Gilbert damping, and the third is the contribution of the spin-torque. The spin-torque can act as damping or anti-damping, depending on the relative effects of the total field and the applied current[7]. Note that in our coordinate system, the magnetization of the pinned layer is parallel to $-\vec{u}_x$ (negative field favors the P alignment).

We solved Eq. (1) following Grollier et al's method[27]. After projection of Eq. (1) on the x, y, z axes, and considering that the magnetization of the free layer is close to either the P or the AP state ($m_x = \mp 1; \dot{m}_x = 0$), the following stability conditions can be deduced for the two orientations of the free layer relative to the pinned layer:

1) the P state is unstable when:

$$I < \frac{2e}{\hbar} \frac{\alpha M_s At}{g(0)} \left( -\frac{H_d}{2} + H_{ms} - H_k + H \right) \quad , \quad \text{if } H < -H_{ms}+H_c; \quad (2)$$

$$I < \frac{2e}{\hbar} \frac{\alpha M_s At}{g(0)} \left( -\frac{H_d}{2} + H_{ms} - H_k + H \right) + \frac{2e}{\hbar} \frac{M_s At}{g(0)} \sqrt{(H_{ms} - H_k + H)(H_d - H_{ms} + H_k - H)} ,$$

if $H < -H_{ms}+H_c$;

2) the AP state becomes unstable for:

$$I > \frac{2e}{\hbar} \frac{\alpha M_s At}{g(\pi)} \left( \frac{H_d}{2} + H_{ms} + H_k + H \right) - \frac{2e}{\hbar} \frac{M_s At}{g(\pi)} \sqrt{-(H_{ms} + H_k + H)(H_d + H_{ms} + H_k + H)} ,$$

if $H < -H_{ms}-H_c$;

$$I > \frac{2e}{\hbar} \frac{\alpha M_s At}{g(\pi)} \left( \frac{H_d}{2} + H_{ms} + H_k + H \right) \quad , \quad \text{if } H < -H_{ms}-H_c;$$
(3)

In order to fit our phase diagram, we have used these formulas to determine the critical lines characterizing the currents at which magnetic switching is observed (Fig. 8). A reasonably good agreement with the experimental data was obtained for the following parameters: $H_c$ = 91 Oe, $H_{ms}$ = 48 Oe, $H_d$ = 16000 Oe, $\alpha$ = 0.006, $g(0)$ = 0.246 and $g(\pi)$ = 0.526. $g(0)$ and $g(\pi)$ were taken as fitting parameters, but the values we found are quite close to those calculated by Stiles and Zangwill[26] for Co/Cu/Co trilayers, and could be used to fit phase diagrams for several samples. The values considered for $H_d$ and $\alpha$ are typical for such structures.

As predicted by Slonczewski's ballistic model (1996)[2], there is a strong asymmetry $I_c^{AP-P} > I_c^{P-AP}$ for H = -51 Oe (when the applied field compensates the magnetostatic interaction from the pinned layer). However, when using Slonczewski's formula for $g(\theta)$ as a function of the polarization of the current, one would expect a much higher $I_c^{AP-P}/I_c^{P-AP}$ ratio then the one we measured.

Above the blue dotted line, the P state is stable; the AP state is stable under the brown dotted line. Consequently, five regions can be distinguished: one region where both states are stable (in the center of the diagram), one region were only the P state is stable, one region were only the AP state is stable and two regions where neither state is stable (high positive/negative fields). In these latter regions, the current generates steady excitations in the free layer. In the macrospin model[27], these steady excitations are identified as a steady precession of the free layer magnetization. This has been recently demonstrated experimentaly[21].

This simple macrospin model predicts correctly several features of the experimental phase-diagram:
1) the general shape of the phase diagram, as well as the existence of four types of regions (P stable, AP stable, both P and AP stable and both P and AP unstable, i.e. precession region);
2) the values of the applied field for which the border lines change from a linear dependence to a parabolic one (H = 43 Oe for P, H = -139 Oe for AP stable), as well as the linear dependence of the instability current for P when H < 43 Oe and for AP when H > -139 Oe, and the parabolic(-like) dependence elsewhere;
3) the values of $I_c^{AP-P}$ and $I_c^{P-AP}$ for H = 0 Oe;
4) the slope of $I_c^{P-AP}$ as a function of H for –210 Oe < H < 43 Oe.

The main discrepancies between the model and the experimental results are:
1) the slope of $I_c^{AP-P}$ as a function of H for H > -139 Oe;
2) the curvature of $I^{end}$ for H < -139 Oe;
3) the linear dependence of the instability current for P, when H < -200 Oe.

The disagreements between the theoretical and the experimental limits of the steady precession region can be at least partially explained by the fact that the model does not consider the influence of the Oersted field generated by the current. This field can be quite important in this region (about 150 Oe for an applied current of 4mA), favoring the formation of a vortex distortion, not taken into consideration by the macrospin model. Several micromagnetic studies have underlined the importance of the Oersted field in the investigation of current induced magnetization switching[9,30]. It was even suggested that in Co/Cu/Co circular nanopillars with a diameter of 130nm (similar to our samples), the field induced by the current plays a crucial role in promoting the switching[30]. Even for samples with important shape anisotropy, several features of the phase diagrams cannot be explained without taking into account the Oersted field[9].

The formation of a vortex distortion in the free layer when increasing the applied (positive and negative) currents also yields a decrease in the magnetoresistance amplitude, but other effects intrinsic to the spin transfer could also contribute. First, under certain conditions, the spin polarized current can induce very fast precession states of the magnetization of the free layer along out-of-plane orbits[9, 10, 21, 22]. Second, different micromagnetic[9] and experimental[12,31] studies have shown that spin transfer can cause telegraph noise either between the P and AP states (inside the coercivity region) or between almost P and AP states corresponding to different precession orbits, even at 0K. The dwell time of such telegraph noise is much shorter (of the order of nanoseconds) than the characteristic time of our experiment (several milliseconds), which means the resistance we measure is statically averaged over this interval. The decrease of the magnetoresistance in high currents is probably the conjoint consequence of all these phenomena, and only high resolution time-resolved and frequency-dependent measurements could shade more light on this point.

The theoretical critical lines on the phase diagram are calculated at T = 0K. It has been shown that thermal effects reduce the critical currents and the switching fields in the coercivity region and cause a rounding of the phase diagram and an increase of the slope of the critical lines in the coercivity region[29]. It is difficult to treat the thermal effects quantitatively for two different reasons: first, for positive currents the temperature of the sample increases very rapidly with the current, while for negative currents the temperature of the sample is approximately constant; second, an Arrhenius-type treatment would not necessarily be appropriate in this case, since the dwell time of the telegraph noise caused by the spin transfer is of the order of the attempt time used in the Arrhenius law of thermal activation ($\tau_0 \sim 1$ns).

From a general comparison of our phase diagram with the theory, we observe that the macrospin model fits the P-AP transition (which occurs mostly for negative currents) better than the AP-P one (observe most of the times for positive currents). Such behavior has been observed earlier in simpler structures. In our samples, the agreement between theory and experiment is expected to be better at negative currents, which are less affected by heating effects.

The effect of the finite temperature on the switching fields is taken into account by using the measured room temperature coercivity of the sample instead of the 0K anisotropy field in the formulas for the critical lines; as a consequence, we find that the experimental critical lines and the theoretical fits change slopes for the same values of the field.

It is important to note that within this simple model, it is only possible to calculate the values of the current where a given state becomes unstable. As it was often commented in the literature, it does not necessarily follow that the free layer actually switches to the opposite orientation. Furthermore, as it was mentioned previously[27], it is somewhat difficult to identify the exact beginning and end of the reversible transitions on the curves, which could also result into discrepancies between experiment and theory. Moreover, it has been shown that the macrospin model is a poor approximation for describing the magnetic dynamics of the free layer during the current induced magnetization switching. Indeed, the dynamics is most often very chaotic during before and during the reversal[9]. Nevertheless, the simple theory described above offers a satisfactory semi-quantitative comprehension of the general features of the I-H phase-diagram.

The macrospin model also fails to explain the presence of the gradual transition towards the AP state at positive current, (small) positive applied field range, as well as the strong telegraph noise measured under the same conditions. Indeed, in our configuration, positive current favors the parallel alignment between the magnetizations of the two layers, and the AP-P transition induced by spin transfer occurs for small positive values of the current. Increasing the current should stabilize the P state. For example, for an applied field of 19 Oe, both the magnetostatic field from the pinned and the external field favor the AP orientation of the two layers. Starting with the sample in the AP state, when sweeping the current from negative to positive values, the $I_c^{AP-P}$ spin-transfer induced transition occurs at $I_c^{AP-P}$ =3.1mA. At this value, the spin-torque effect becomes stronger than that of the resultant field acting on the magnetization of the free layer. Since, for the positive current, the electrons flow from the pinned to the free layer, increasing the current should not change the state of the sample. However, at 5mA, the free layer starts to relax back to a higher resistance state, as seen in Fig. 4. This second transition cannot be explained by the formation of a vortex induced by the Oersted field of the current, since the resistance at 8mA is closer to the resistance of the AP state than to that in a vortex configuration. Moreover, this second transition shifts towards lower currents when the applied field is increased. This saturation effect could be interpreted as a loss of spin-polarization due to the high temperature increase for positive currents (see discussion of Fig.5 in paragraph results). Since the temperature significantly increases with the current, the polarization is probably reduced due to enhanced spin-flip by magnon scattering. This effect can be particularly pronounced in laminated structures in which each individual magnetic layer is quite thin (~1nm). As a result, when the effect of the total field acting on the magnetization of the free layer becomes stronger than the spin transfer, the system relaxes back to the antiparallel state. On the other hand, in the negative currents the sample heats much less than in the positive currents, so that the depolarization effect does not take place. Consequently, this second transition does not exist at negative currents.

Alternatively, another explanation could be proposed for the depolarization of the spin current at positive current in terms of effective magnetic temperature. It has been recently shown that when spin transfer and field have opposite effects, the spin torque excites incoherent magnetic excitations above a certain current threshold $I_{threshold}$[9]. Of course, these excitations cannot be described in a single domain model. However, they can be described by a concept of effective magnetic temperature $T_m$. The latter increases quite significantly above $I_{threshold}$[9] which may create the same type of depolarization effect than the increase of temperature due to Joule and Peltier effects.

Telegraph noise appears when the spin transfer torque and the applied field have opposite effects that almost compensate each other. In this case, for H =1 Oe, $H_{res}$ = 49 Oe (considering a uniform magnetostatic field from the pinned layer), telegraph noise is observed for all current values between 2 and 8mA, i.e. on all the range of currents where the sample should be in the P state; this means that on all the range the current and the field approximately compensate, which implies that the spin transfer is not increasing with the applied current density. No telegraph noise is observed between –8 and 2mA, where the sample is in the AP state (both the field and the negative applied current favor the AP state and the system is stable). Increasing the applied field reduces the noise, since the current cannot compensate this stronger field, but introduces the second transition (relaxation towards the AP state at high positive current). Increasing the field causes a shift of this transition towards lower values of the current, as well as a further decrease of the telegraph noise.

Finally, as we have argued before[19], we emphasis that we observed spin-transfer induced effects in these samples, although the AP1 polarizing layer (4.4nm) is much thinner than the hard magnetic layer in commonly used samples in which spin transfer effects were investigated before. The thickness of the AP1 layer is even smaller than the spin diffusion length in bulk $Co_{50}Fe_{50}$ (6±1nm)[28]. Because of the lamination of the pinned layer, as discussed in the introduction, the effective spin diffusion length in the laminated stack is reduced to 1.2±0.1nm due to increased interfacial scattering and higher density of thermally activated magnetic fluctuations[28]. Consequently, a significant current polarization can build up in this layer, which explains the large measured spin transfer effects.

**Conclusion**

We have demonstrated that large spin-transfer induced effects, in particular current induced magnetization switching, can be observed in complex spin-valve structures developed for CPP-magnetic heads. The macrospin model can reasonably well account for the experimental results. A spin-transfer 'saturation' effect was observed for the positive direction of the current; we interpret it as a spin-depolarization effect at large positive currents resulting from the asymmetric heating of the pillars (Peltier effect) and possibly to the onset of incoherent excitations due to the spin transfer.

**Acknowledgements**

The authors would like to thank Pr. A.Vedyaev, Pr N.Ryzhanova and Dr. N. Strelkov for fruitful discussions, and Dr. U. Ebels and Dr. D. Stanescu for helping with the experimental set-up. This work was partially supported by the IST project NEXT (IST-37334) and the RMNT project MAGMEM II.

**Figure Captions**

**FIG. 1.** The structure of the samples includes a laminated CoFe free layer and a laminated synthetic pinned layer. The pillars have a "square" section with a lateral size of 130 nm. The magnetization of the AP1 (polarizing) layer is oriented along the $-\mathbf{u_x}$ direction; the current is flowing along the $\mathbf{u_z}$ direction.

**FIG. 2. a:** Magnetoresistance curve measured with a sense current I = – 0.400 mA; the blue curve is measured for increasing fields, the pink one for decreasing fields. Negative fields are oriented along the magnetization of the pinned layer (favor the P state). **b:** Resistance versus current characteristics for H = - 4 Oe. The green curve is measured for increasing currents, the pink one corresponds to decreasing currents. For the positive current, the electrons flow from the pinned to the free layer (see fig.1), thus favoring the P alignment.

**FIG. 3.** Resistance versus current characteristics for increasing values of the negative applied field.

**FIG. 4.** Resistance versus current characteristics for increasing values of the positive applied field. The thin grey curve (H = 19, 25, 37 Oe) is measured at 600 Oe, when the sample is in the AP state.

**FIG. 5.** Temperature variation of $R^{AP}$ and $R^P$ normalized to the respective values at room temperature, measured for a second sample with an identical structure.

**FIG. 6.** Magnetoresistance curves for increasing values of the negative sense current.

**FIG. 7.** Magnetoresistance curves for increasing values of the positive sense current.

**FIG. 8.** The phase diagram obtained from the resistance versus current curves for constant applied (negative) magnetic fields, and from the magnetoresistance curves measured with different (negative and positive) sense currents. We have plotted the switching currents ($I_c^{AP-P}$ and $I_c^{P-AP}$) or the switching fields ($H^{AP-P}$ and $H^{P-AP}$) if the loop was hysteretic, and the beginning and the end of the transitions ($I^{start}$ and $I^{end}$, or $H^{start}$ and $H^{end}$), if they were reversible. The dotted lines represent the theoretical fit.

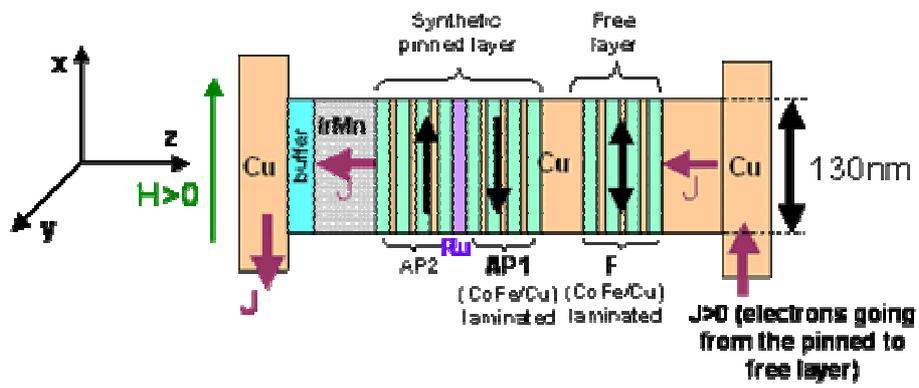

FIG.1

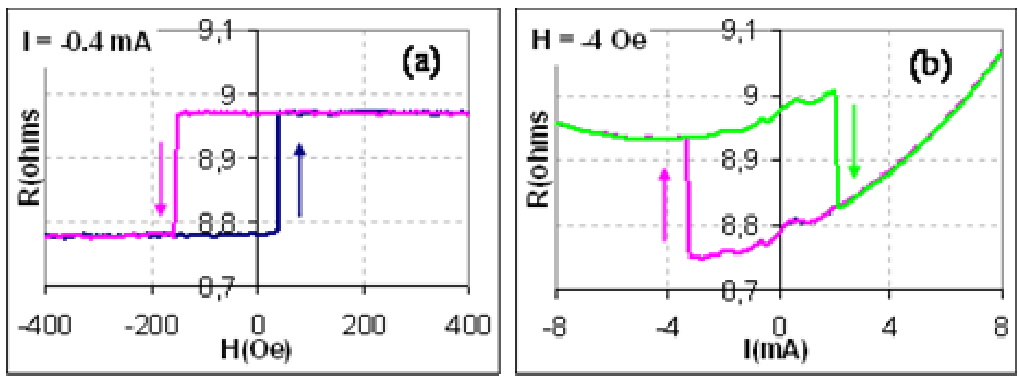

FIG.2

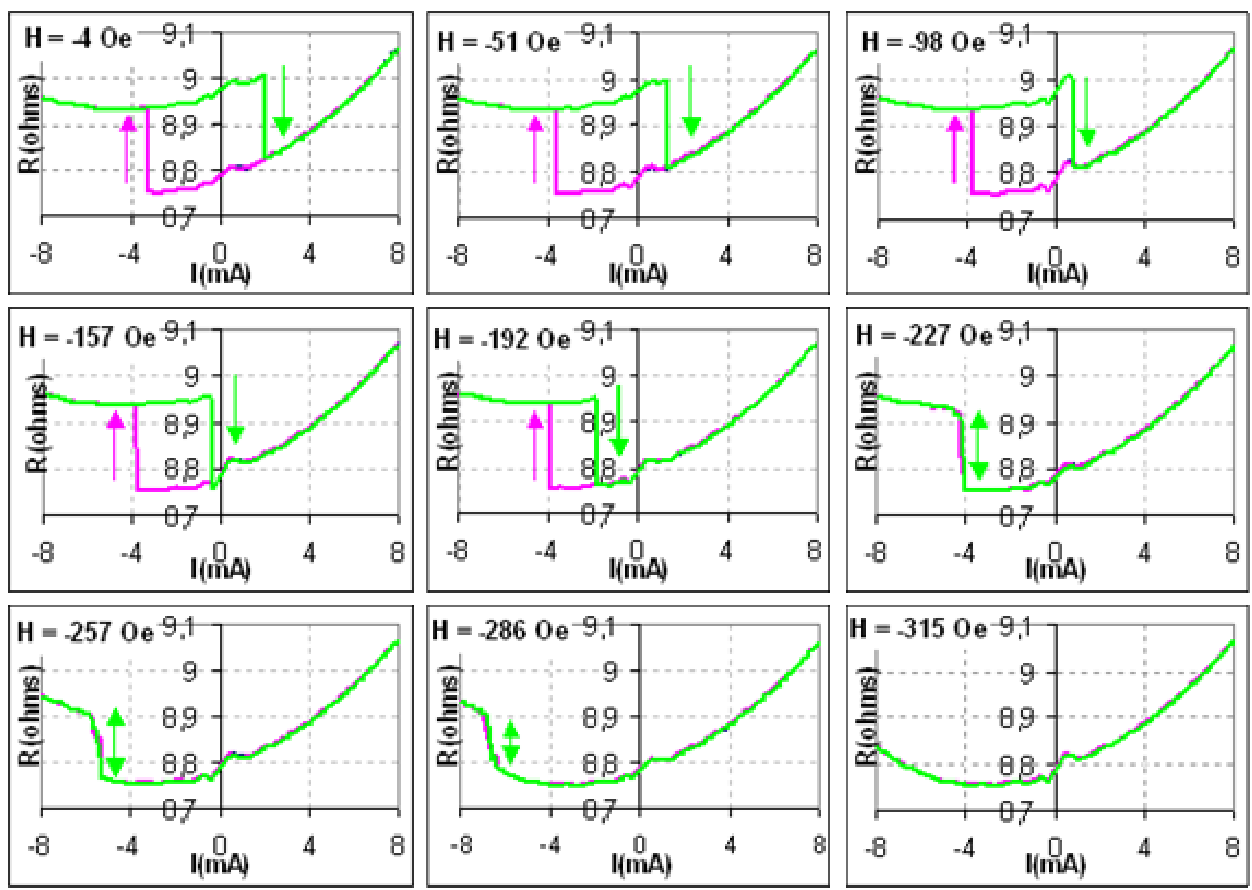

FIG. 3

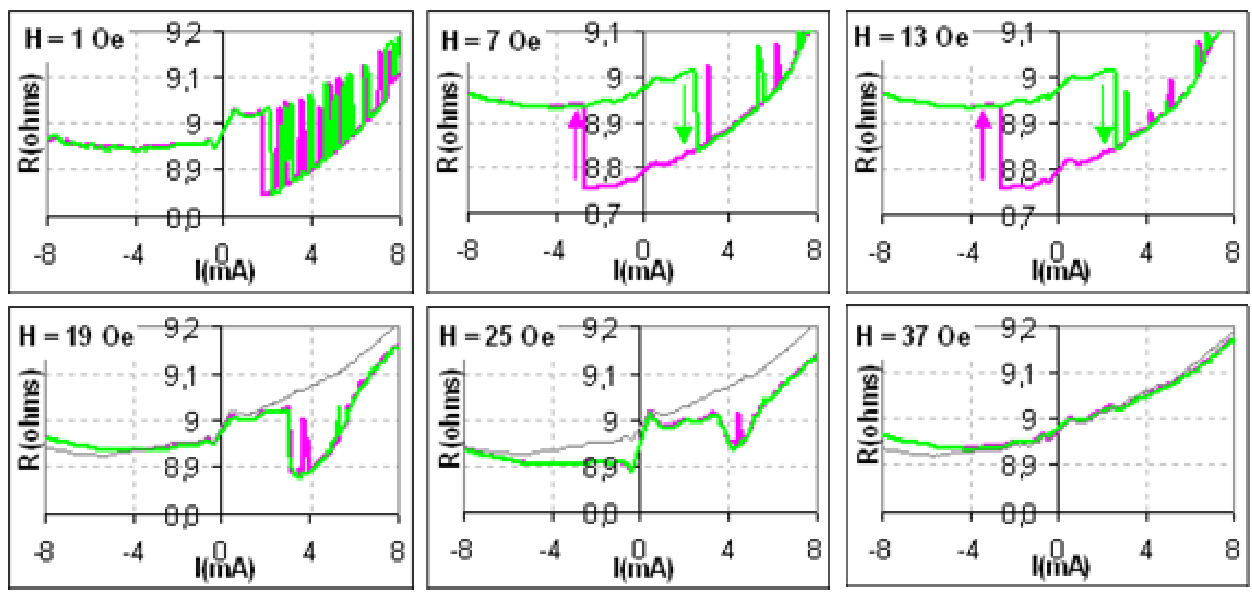

FIG. 4

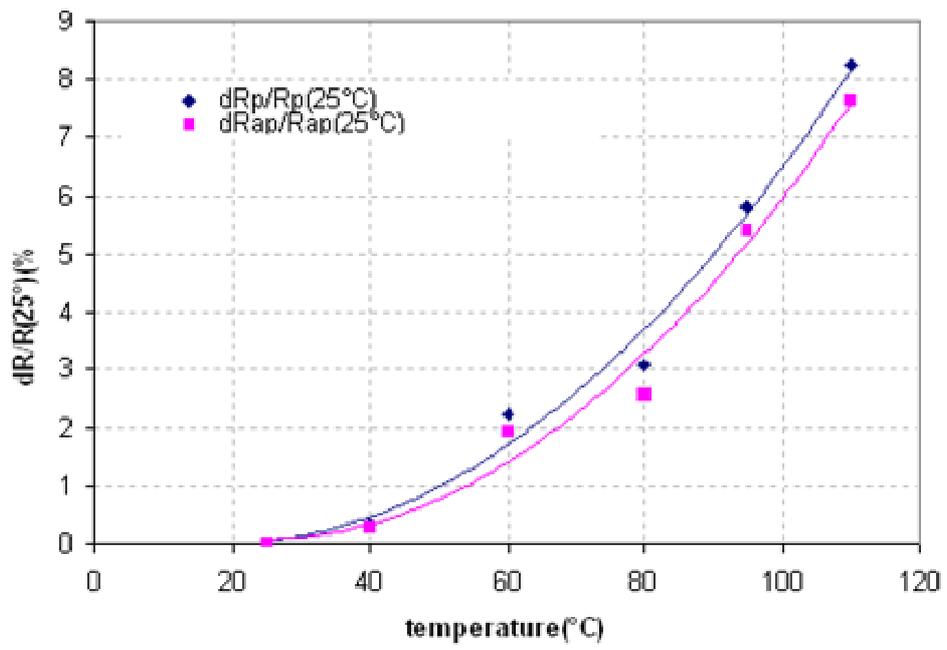

FIG 5

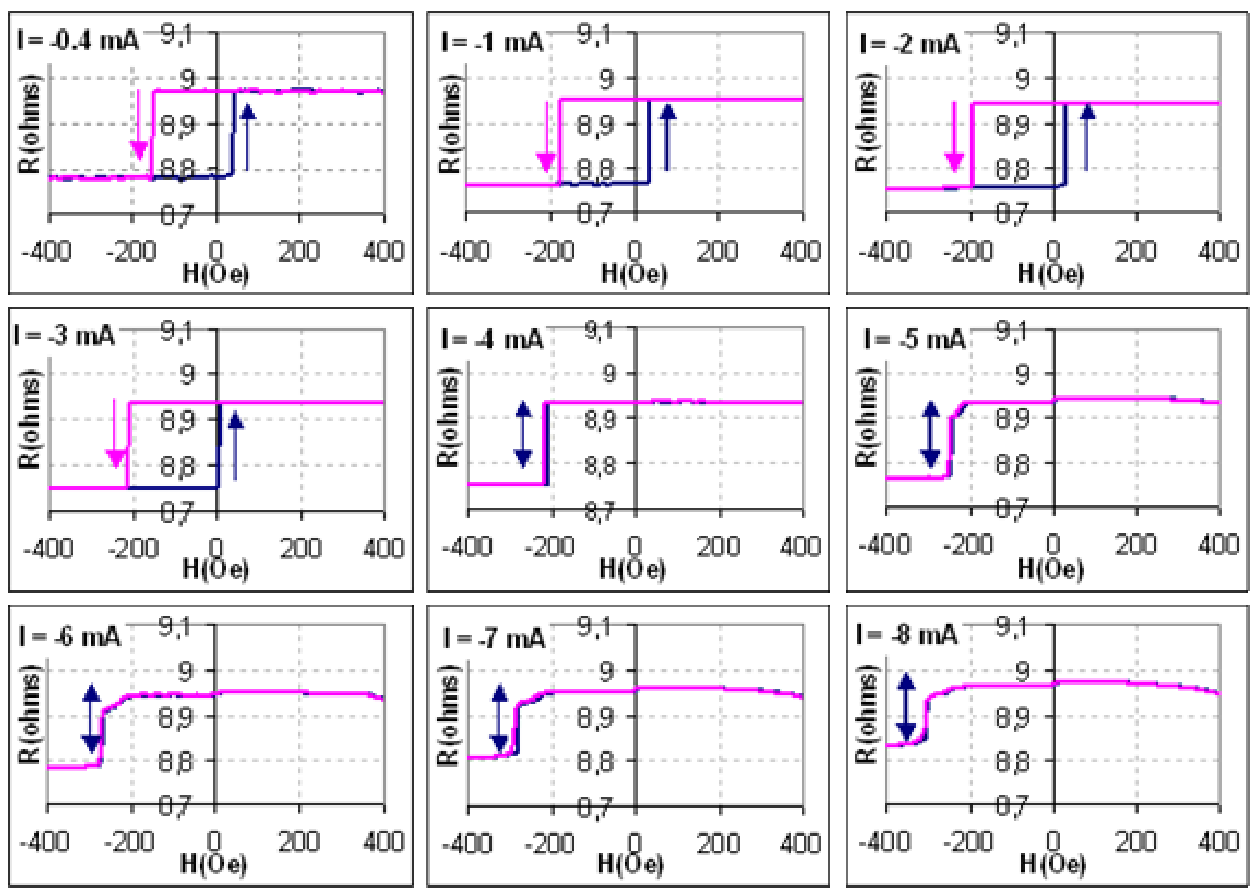

FIG.6

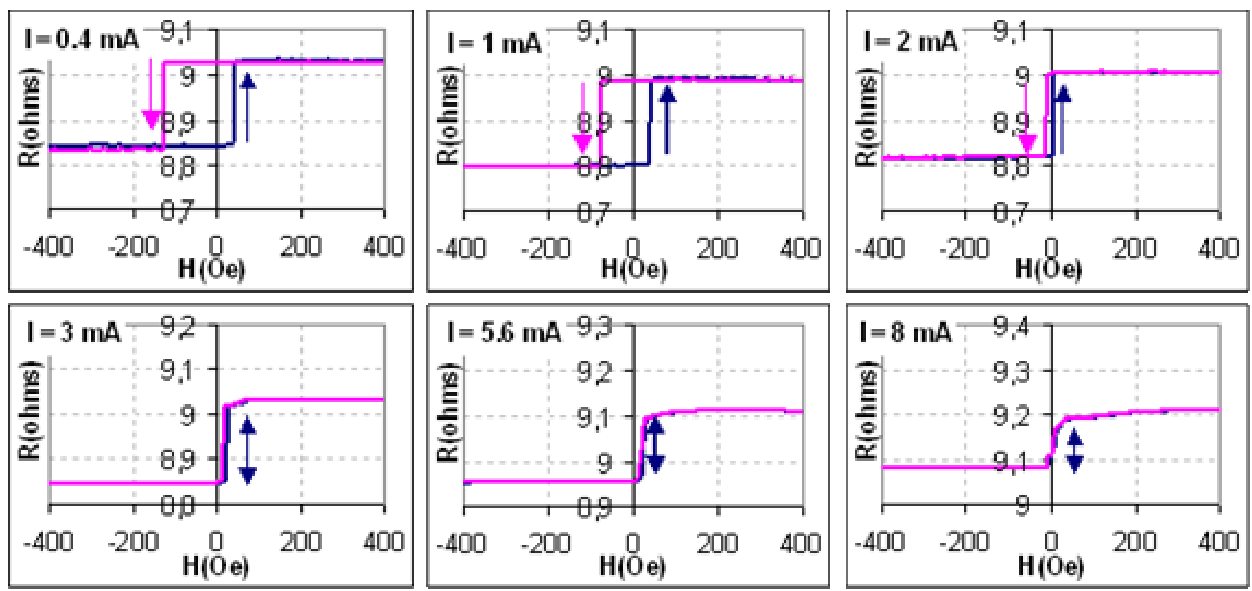

FIG.7

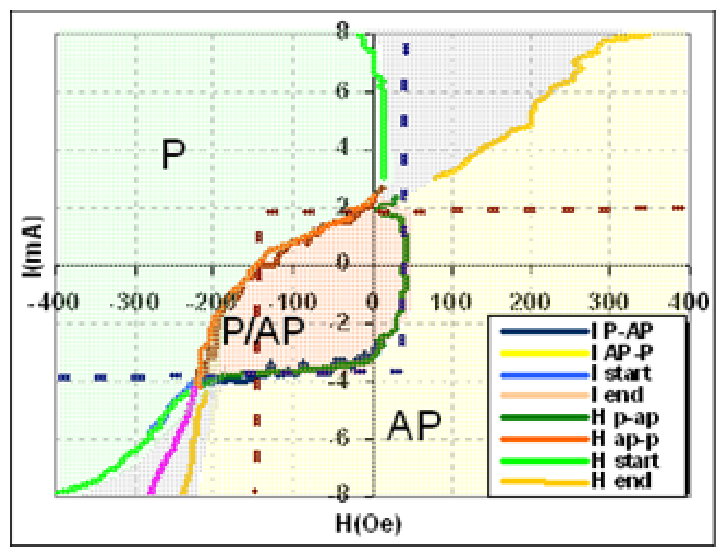

FIG. 8